%% file: main.tex
\documentclass[onecolumn,showkeys,reprint,superscriptaddress,nofootinbib,amsmath,amsfonts,amssymb,aps,floatfix,]{revtex4-1}
\usepackage{amsmath,amsfonts,amssymb}
\usepackage{subfigure}
\usepackage{graphics,graphicx}
\usepackage{xcolor}
\usepackage{url} 
\usepackage{hyperref}
\usepackage[a4paper, total={7in, 10in}]{geometry}

\begin{document}

\title{Temporal clustering of social interactions trades-off disease spreading and knowledge diffusion}

\author{Giulia Cencetti}
\affiliation{Fondazione Bruno Kessler, Trento, Italy}
\author{Lorenzo Lucchini}
\affiliation{DONDENA and BIDSA research centres - Bocconi University, Milan, Italy.}
\author{Gabriele Santin}
\affiliation{Fondazione Bruno Kessler, Trento, Italy}
\author{Federico Battiston}
\affiliation{Central European University, Vienna, Austria}
\author{Esteban Moro}
\affiliation{Massachusetts Institute of Technology, Cambridge, MA, United States}
\author{Alex Pentland}
\affiliation{Massachusetts Institute of Technology, Cambridge, MA, United States}
\author{Bruno Lepri}
\affiliation{Fondazione Bruno Kessler, Trento, Italy}

\begin{abstract}
Non-pharmaceutical measures such as preventive quarantines, remote working, school
and workplace closures, lockdowns, etc. have shown effectivenness from an epidemic control perspective; however they have also significant negative consequences on social life and relationships, work routines, and community engagement. In particular, complex ideas, work and school collaborations, innovative discoveries, and resilient norms formation and maintenance, which often require face-to-face interactions of two or more parties to be developed and synergically coordinated, are particularly affected.
In this study, we propose an alternative hybrid solution that balances the slowdown of epidemic diffusion with the preservation of face-to-face interactions. 
Our approach involves a two-step partitioning of the population. First, we tune the level of node clustering, creating ``social bubbles" with increased contacts within each bubble and fewer outside, while maintaining the average number of contacts in each network. Second, we tune the level of temporal clustering by pairing, for a certain time interval, nodes from specific social bubbles.
% We compare the node-temporal clustering approach with the conventional preventive quarantine strategy, which serves as the benchmark for disease containment, maintaining the same dynamical properties for the two diffusion processes.
Our results demonstrate that a hybrid approach can achieve better trade-offs between epidemic control and complex knowledge diffusion. The versatility of our model enables tuning and refining clustering levels to optimally achieve the desired trade-off, based on the potentially changing characteristics of a disease or knowledge diffusion process.

\end{abstract}

\maketitle

\section{Introduction}

\begin{figure*}
\centering
\includegraphics[width=0.5\textwidth]{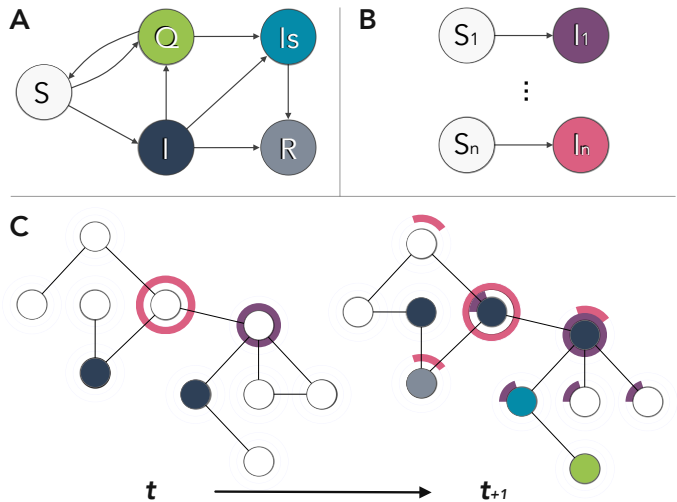}
\caption{\emph{Schematic representation of the combined model dynamics}. A) The epidemic spread is modelled through a modified SIR compartmental model framework. ``$S$'' are individuals susceptible to the infection, ``$Q$'' are quarantined individuals, ``$I$'' are infected individuals, ``$Is$'' are isolated individuals, and ``$R$'' are recovered individuals. B) The information dynamics follows a standard multistrain SI compartmental model. With ``$S_i$'' representing agents susceptible to information ``$i$'', and ``$I_i$'' representing agents who acquired information ``$i$''. C) Schematically reproduce the main aspects of the two dynamics over a network of agents. Agents are represented with three concentric circles. The colour of the inner circle represents the epidemiological compartment the agent is in; the colour of the outer circles represents the progression of the information diffusion model from time ``$t$'' to time ``$t+1$''.}
\label{fig1}
\end{figure*}

The recent experience with COVID-19 has made us all aware of epidemics, their possible appearance and their likelihood to overturn our lives all of a sudden. The COVID-19 pandemic, as of June 2023, has caused almost 800 million cases (among which 7 million deaths) around the world\footnote{World Health Organization https://covid19.who.int}. Many containment measures and non-pharmaceutical interventions (NPIs) have been put in place before vaccines became available: lockdowns, preventive quarantines, masks, physical distancing, school and workplace closures, remote working, etc. \cite{perra2021non,hsiang2020effect,dehning2020inferring,brauner2021inferring,soltesz2020effect}.
%ludecke2020protective,yang2021despite,zhang2020changes,flaxman2020estimating
All these measures have considerably impacted people's lives, social relationships, work, and economy \cite{wbg2022helping,yabe2023behavioral,hunter2021effect,lucchini2023socioeconomic,lucchini2021living}. 
Indeed, if these measures can represent possible solutions to reduce disease spreading, the other side of the coin is that they imply a severe slowdown, or even interruption, of all the social exchanges and face-to-face interactions which prove fundamental for the proper functioning of a society and, specifically, for fruitful collaborative interactions.

Moreover, despite many collaborations, intimate relationships, and, more in general, information and knowledge sharing can nowadays easily travel through the Internet, the limits of only remote interactions have become clear \cite{seitz2020pandemic,van2020daily,myers2020unequal}. %\cite{balanza2021assessment}. 
Several studies, indeed, have highlighted the importance of in-person interactions for physical, psychological and social wellbeing~\cite{probst2020changes,jeste2020changes,khan2020covibesity,sutin2020change}. In particular, observational, interview-based, and questionnaire-based studies have found that in workplaces face-to-face interactions are associated with increased trust and improved communication among employees, efficient problem solving and a positive effect on the overall organization knowledge diffusion, innovation ability, and performance~\cite{whittaker1994,teasley2000,kirkman2004,salis2010}. Similar results have been also found using wearable sensors to study face-to-face interactions and their effect on productivity, performance and complex tasks' completion~\cite{wu2008,pentland2012,wilson2013}.\\
%(\cite{lambrese2020helping}). 
%colizzi2020psychosocial,murphy2021challenges,rogers2020behavioral
% Moreover, despite many collaborations, information transfer, and even emotional connections can nowadays largely travel through the Internet, the limits of only remote interactions are now clear and highlight the importance of in-person interactions for physical, psychological and social wellness. 
For this reason it is crucial, along with the epidemic spreading reduction, to maintain as many physical interactions as possible. These two objectives are not easily pursued simultaneously. 
In this work, we will try to find a trade-off between them by working on the design of the network of social interactions, a useful exercise to find alternative solutions allowing us to cope with a fast-spreading disease. Tackling this problem within a coupled-dynamics framework can nourish new perspectives for the future management of epidemics and health emergencies.

% modify network
Since the behaviour in time of a spreading process on a network is heavily conditioned by the network topology, several studies tried to regulate spreading by acting on the network structure~\cite{shirley2005impacts,ikeda2010cascade,hackett2011cascades,ansari2021moving}. 
%bubbles
In this work, inspired by \cite{block2020social}, we explore the ``social bubbles" strategy, which implies partitioning the society into communities where each individual can physically interact at will inside the bubble but in a controlled amount (or not at all) outside. In this way, people can maintain a normal amount of face-to-face interactions but restrict them to a set of people who interact exclusively in the same group. For example, in a workplace this could imply restricting face-to-face interactions only among the members of the same team or department, while in a school and in a university campus only within a classroom or a dorm. This strategy has been proposed and largely discussed, and many numerical experiments have been performed to assess the effect of social bubbles on real populations~\cite{willem2021impact} and in specific contexts like schools~\cite{gemmetto2014mitigation,gauvin2015revealing,mcgee2021model,leoni2022measuring} and workplaces~\cite{valdano2021reorganization}, always showing important advantages in reducing contagions. 
% communities and spreading on synthetic networks
In order to evaluate the effect of the community structure it is useful to consider synthetic networks where we can tune the modularity, i.e. the strength of division into modules or bubbles, and observe how this affects the spreading~\cite{gleeson2008cascades,nematzadeh2014optimal,peng2020network}.\\
%In particular, our main focus is to regulate the disease spreading while in parallel trying to minimize the social damage~\cite{antelmi2021modeling,meidan2021alternating}. To do this we will introduce also a spreading of knowledge or social behavior~\cite{cowan2004network,ok2014maximizing,lambiotte2009communities,chung2014generalized}, which can be combined with the disease spreading~\cite{she2022networked,peng2021multilayer}. \giulia{Questa frase si può anche cancellare e spostare le citazioni sotto quando si introduce la knowledge}
In this work, we explore the effect of the social bubbles' reorganization of a network of proximity interactions. The specific goal of this study is to find an optimal topological network structure that minimizes the number of infected individuals (and in particular the number of simultaneously infected individuals) in order to avoid burdening hospital intensive care units (ICUs), and, at the same time, minimizing the social deterioration due to restrictions~\cite{antelmi2021modeling,meidan2021alternating}. 
For this reason, we consider two different and non-interacting spreading processes: one regarding a disease, and one regarding the diffusion of knowledge or of a social behaviour~\cite{cowan2004network,ok2014maximizing,lambiotte2009communities,chung2014generalized,she2022networked,peng2021multilayer}. The first one is represented by a simple contagion model, a SIR (Susceptible, Infected, or Recovered) compartmental model inspired by \cite{anderson1991infectious,pastor2015epidemic}, while the second one is governed by a complex contagion approach, the threshold model with memory~\cite{dodds2004universal}.
The two processes take place simultaneously on an artificial population where individuals are connected via a temporal network, i.e. a set of pairwise links that appear and disappear in time. The way these links are distributed among nodes heavily affects the temporal evolution of the two spreading processes. 
The effect of social bubbles, which in networks is reflected by nodes clustering, is investigated for different sizes of the groups (e.g. different sizes of teams or departments in a workplace, different sizes of classrooms at school or university, etc.) and different levels of modularity (i.e. the strength of network partition, represented by the connectivity inside each bubble with respect to the admitted contacts between bubbles). Importantly, all these different networks have the same average number of links, such that our analysis is not influenced by the number of connections (a parameter that clearly has a role in fastening all kinds of spreading processes).

We show that it is possible to find a trade-off between minimizing the timescale of the knowledge diffusion and the number of simultaneous infected individuals, two competing objectives. 
The effect of social bubbles (without preventive quarantines) is compared with the effect of quarantines (on a network that is not organized in bubbles).\\
We will show that, even if the quarantines are more effective in containing the number of infected, they do not allow knowledge diffusion until most of the population is recovered. In contrast, with the bubbles strategy it is instead possible to share knowledge in the network since the beginning of the simulation, and simultaneously maintain the number of infected below a critical level (generally higher but comparable with the case of quarantines). The bubbles strategy, therefore, allows social processes to coexist with an epidemic. Additionally, in agreement with \cite{nematzadeh2014optimal,peng2020network} we also find an optimal value of modularity for information diffusion, revealing a non-monotonic relation between knowledge diffusion and network structure.

%%%%%%%%%%%%%%%%%%%%%%%%%%%%%%%%%%%%%%%%%%%
%%%%%%%%%%%%%%%%% RESULTS %%%%%%%%%%%%%%%%%
\section{Results}

% disease + knowledge
Combining simple and complex contagion allows in general to find strategies that take into account both the epidemic threats and the socio-economical issues deriving from prolonged isolation periods, societal fragmentation, and, potentially, segregation. Several works exist studying the interplay between different spreading processes~\cite{she2022networked,peng2021multilayer,lucas2023simplicially}. These works however consider that the two processes mutually interact or one strongly affects the other one, while in our work they are considered two parallel processes (knowledge does not affect disease, and disease affects knowledge only indirectly, via isolations and quarantines). The focus of this work is indeed on the network structure and how this can be set so as to regulate different spreading processes taking place on it.  We hence chose to consider a setting where the interaction between the two processes is minimal, thus avoiding to insert additional effects with the risk of not being able to understand from what they are generated, and hence to complicate the results' interpretation.

The disease starts from one random infected node at time 0. Simple contagion implies that nodes can only be infected if they have an infected neighbour and each contagion event is independent of the other ones. When a connection appears between an infected and a susceptible node, the probability that the susceptible node gets infected is set by $\omega(\tau)$, which depends on the age $\tau$ of the infected node's infection, i.e. the time since it has been in turn infected (see Methods for more details). Infected individuals can be identified and isolated, which means that all their connections are cut and they cannot spread disease or knowledge for a fixed interval of time. This happens with a probability per unit time $\varepsilon_I$, one of the parameters of the model. Infected nodes, whether they are isolated or not, eventually become recovered, hence immune. This is the baseline model that we use to simulate the disease spreading and to investigate social bubbles, but we can also additionally consider the existence of quarantines: in this case, once a node is isolated, its last contacts are traced and preventively quarantined with a probability $\varepsilon_T$, which means cutting the nodes' contacts for a short interval of time without knowing if it is actually infected or not. All infected nodes that are not isolated or quarantined are classified as active infected ones, being free to spread the disease (see Methods for a more detailed description of the disease spreading).\\
In parallel, knowledge spreads across the network. We consider 20 different pieces of knowledge distributed in the network, they can be thought of as 20 different pieces of information or expertise spread among different teams or departments in a workplace. Initially, each piece of knowledge is possessed by only one random node and we assume that the other nodes need multiple exposures in order to acquire them. This is represented by a threshold model with memory~\cite{dodds2004universal}: each susceptible node becomes infected (i.e. it acquires a specific knowledge) only after $K$ interactions with nodes possessing that particular knowledge. Nodes progressively store and accumulate the different pieces of information that they get in touch with and only when the threshold is reached for a particular knowledge it is considered as acquired. This is an SI model: we assume that knowledge cannot be unlearned. See Fig.~\ref{fig1}C for a schematic representation of the two processes: 
%a scheme of the disease compartmental model is depicted in Fig.~\ref{fig1}A and that of the knowledge spreading in Fig.~\ref{fig1}B.
infected nodes (dark blue) infect their neighbours and then become recovered (grey) or isolated (bright blue) and their neighbours can be quarantined (green). In the meanwhile, nodes which possess a particular knowledge pass pieces of it to their neighbours. The neighbours start to collect them and when they obtain the entire set of $K$ pieces that knowledge is acquired. In Fig.~\ref{fig1} only two different spreading pieces of knowledge are represented, the purple and the pink one, while in the numerical experiments we consider 20 pieces of knowledge.\\
We analyze several scenarios in which the two processes can interact and several networks of contacts.
The temporal networks are generated building each layer with the Stochastic Block Model~\cite{holland1983stochastic} where the number of nodes is fixed ($N=680$) and the number of links is fixed on average (400 links). The number of nodes and connections are chosen so as to mimic the proximity interactions' dataset of the Copenhagen Network Study~\cite{Sapiezynski2019}. The population is partitioned in several communities which are strongly connected inside (random temporal connections with a probability $p_{intra}$) and poorly connected between each other (random temporal connections with a probability $p_{inter}$).
Since the number of links is fixed, the values of $p_{intra}$ and $p_{inter}$ are not independent of each other, and increasing $p_{intra}$ implies decreasing $p_{inter}$ and vice-versa. Thus, we rely on a parameter $p = p_{intra}/p_{inter}$ which represnets the network modularity: namely, tuning the value of $p$ allows us to generate more self-contained bubbles (higher $p$) or more interconnected groups less connected inside (lower $p$). By doing this, we can explore the effect of stronger or weaker bubbles as possible ways to reorganize a network without cutting or adding any link (see Methods for a detailed description of network generation). 

\begin{figure*}
\includegraphics[width=\textwidth]{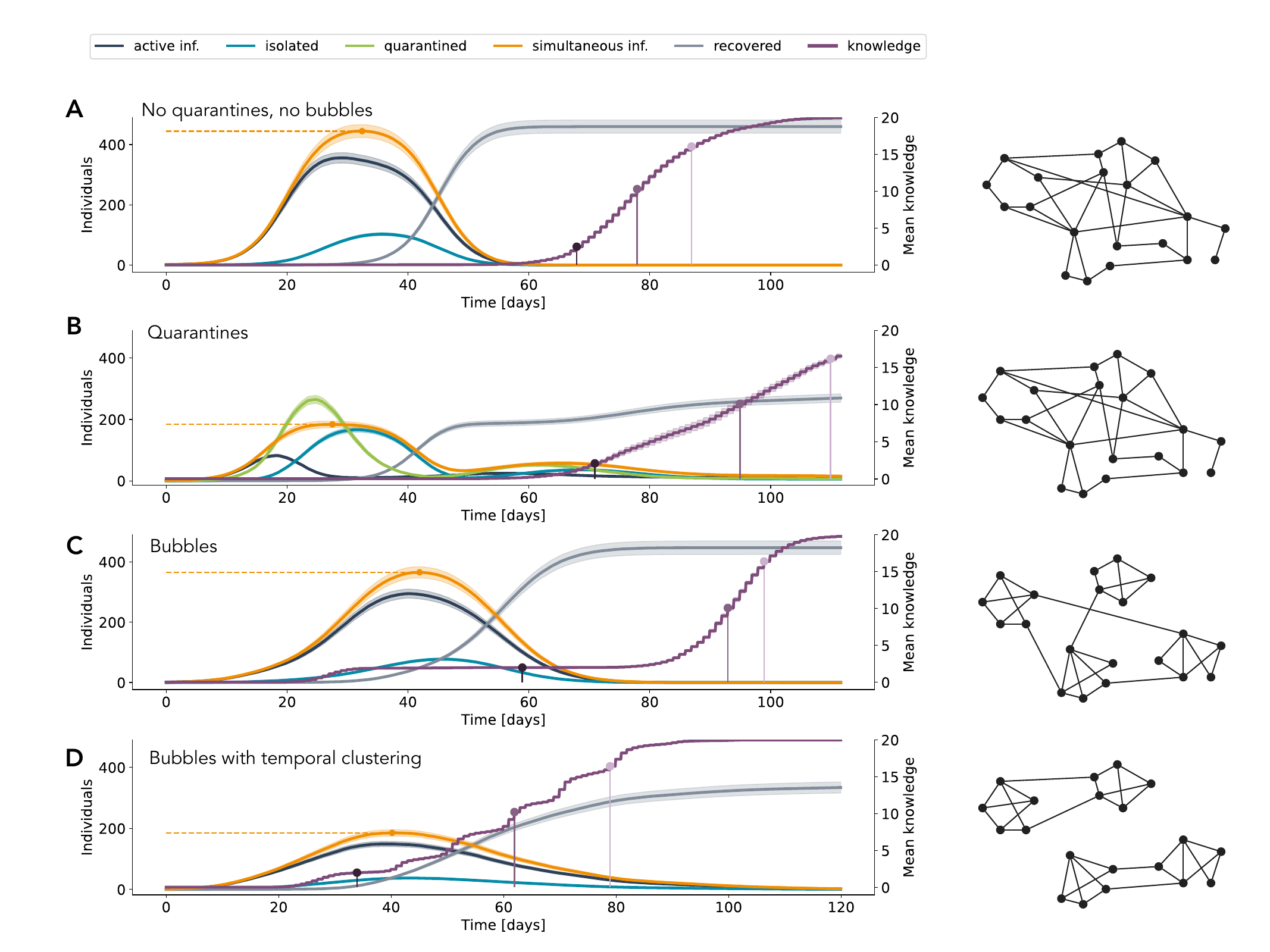}
\caption{Time evolution of disease and knowledge spreading, results of 200 simulations on temporal networks organized in 10 bubbles with 68 nodes. A) $p=5$ (i.e. connections inside and outside bubbles are of the same order, in practice, bubbles do not exist), $\varepsilon_T = 0$ (i.e. no quarantines). B)  $p=5$, $\varepsilon_T = 0.1$ (i.e. quarantines without bubbles). C) $p=199$, $\varepsilon_T = 0$ (i.e. bubbles without quarantines). D) $p=199$, $\varepsilon_T = 0$, plus temporal clustering of 10 days.
The parameter $\varepsilon_I = 0.1$ for all these cases.}
\label{fig_time_evol}
\end{figure*}

\begin{figure*}
\subfigure[]{\includegraphics[width=\textwidth]{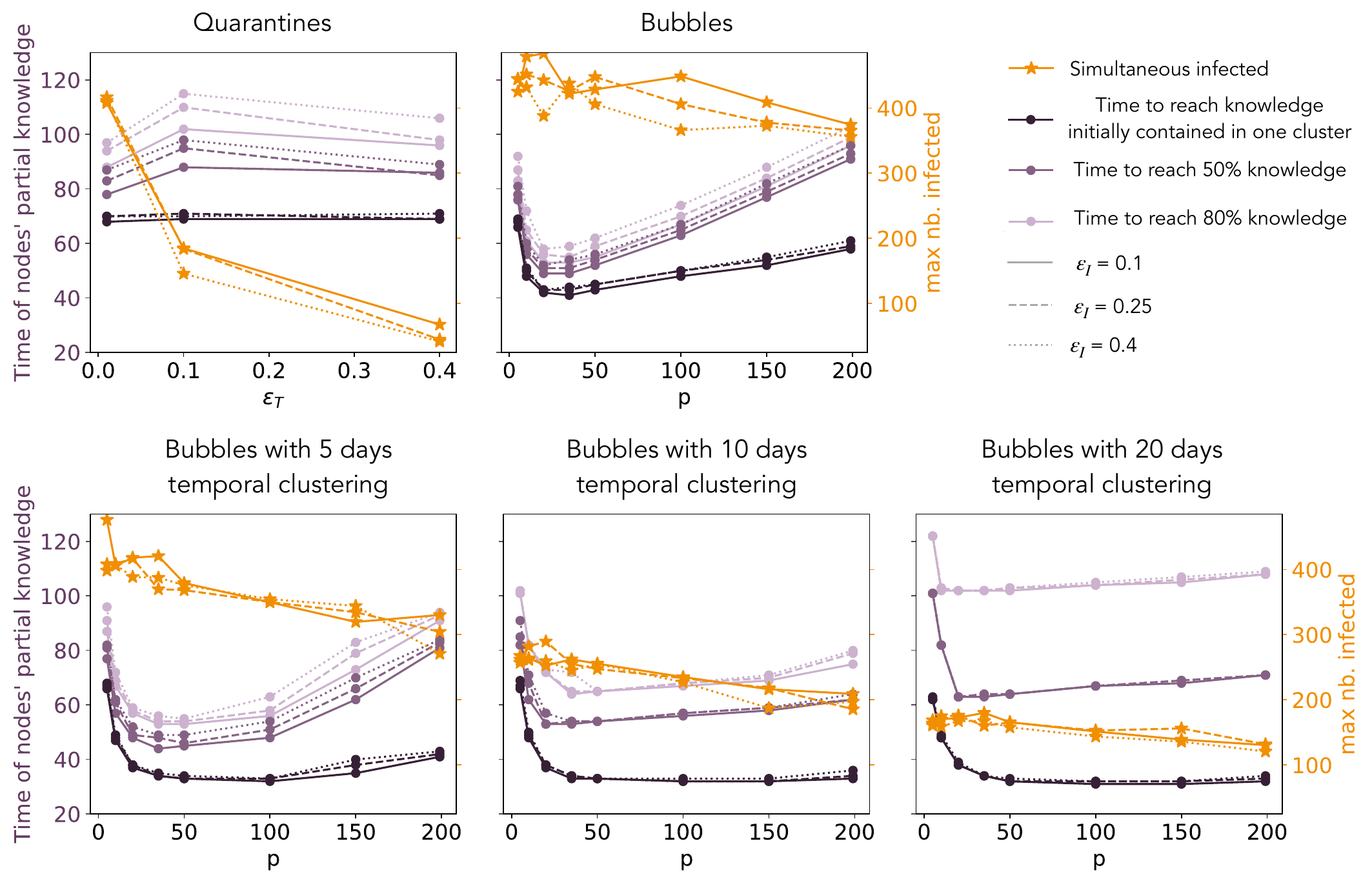}}
\caption{Scenario with 10 bubbles (i.e. ten different teams or departments in an organization). Orange stars represent the maximum number of infected individuals, the other symbols represent the time at which the nodes reach on average 10\%, 50\%, and 80\% of knowledge. The first percentage corresponds to acquiring the entire knowledge initially contained in one bubble from all the nodes. We report results for $\varepsilon_I = 0.1$ (continuous lines), $\varepsilon_I = 0.25$ (dashed lines), $\varepsilon_I = 0.4$ (dotted lines). The quarantine results are reported vs. $\varepsilon_T$, the bubbles results vs. $p$ and for different levels of temporal clustering.}
\label{fig_opt_p_10cl}
\end{figure*}

In Fig.~\ref{fig_time_evol} we depict the evolution in time of both disease and knowledge spreading quantities: the number of active infected (dark blue) individuals, isolated (light blue) individuals, and quarantined (green) individuals, plus the number of simultaneous infected (orange) individuals corresponding to the sum of active infected, isolated and true positive quarantined individuals (i.e. those that are actually infected). The horizontal dashed orange line highlights the maximum number of simultaneous active infected individuals reached during the temporal evolution. Additionally, we plot in purple the evolution of the mean number of different knowledge obtained by the nodes (averaged on all the nodes). We report mean knowledge in the y-axis on the right, which spans from 0 to 20 since we are considering 20 pieces of knowledge spreading in the network (at time 0 each piece of knowledge is only possessed by one node so the mean is $1/N=0.0015$). We highlight three significant steps of knowledge spreading: the time when the average reaches the number of pieces of knowledge initially contained in one cluster, $20/n$ with $n$ being the number of clusters ($n=10$ in the example of Fig.~\ref{fig_time_evol}); the time when it reaches 50\% of total pieces of knowledge (10 in this case); the time when it reaches 80\% of total knowledge (16 in this case). All the reported results are obtained as averages over 200 stochastic simulations.\\
% No quarantines, no bubbles
The first scenario that is depicted (Fig.~\ref{fig_time_evol}A) corresponds to the case without quarantines ($\varepsilon_T=0$) and where the social bubbles' strategy is not at play ($p=5$, meaning that intra-bubbles and inter-bubbles connections are of the same order of magnitude). The only NPI is represented by the isolation of individuals who are identified as infected. This is the worst case: we have a peaked curve of infected individuals and at the maximum peak more than half of the population is simultaneously infected. The average knowledge stays around 0 for the entire time span where people are infected and starts to grow only after the epidemic has been controlled.\\
% Quarantines
Then, we introduce the scenario that serves as a benchmark to compare the social bubbles' strategy: it is the one with quarantines ($\varepsilon_T = 0.1$), still without the bubbles' organization of the network of interactions ($p=5$). Quarantines clearly have an effect on infections number, managing to flatten the curves so as to reduce the number of simultaneously active infected individuals. However, this reduction comes at a great social cost, confining a significative fraction of healthy individuals. In the example reported in Fig.~\ref{fig_time_evol}B, while the maximum average number of simultaneous infected is only %183.8 (
the $27\%$ of the population, the average percentage of population  confined at least once without being infected (collateral confinement) is as high as $30.6\%$ (see Supplementary Material Section \ref{sec_confinement} for collateral confinement in simulations with different parameters). Moreover, we observe that flattening the curve also implies extending the time span of the epidemics and this, in turn, badly affects the possibility of people having face-to-face interactions, thus slowing down knowledge diffusion. In fact, we observe that the purple curve starts to grow only when the other curves are very low. The time needed to acquire knowledge is hence longer than in the previous case. In other words, also in this case knowledge spreading via physical interactions can hardly coexist with an ongoing epidemic.\\
% Bubbles
In Fig.~\ref{fig_time_evol}B we finally introduce the social bubbles' strategy. In this case, no quarantine strategy is put in place (i.e. $\varepsilon_T = 0$) but the network is generated with $p=199$, meaning that the intra-bubbles' interactions are, on average, 199 times more frequent than inter-bubbles' interactions. The number of infected individuals is higher with respect to the previous case: a simple organization of the network of interactions in bubbles is not able to reduce infections as quarantines, even if they are reduced with respect to the case without a bubble structure (Fig.~\ref{fig_time_evol}A). However, we notice that in this case knowledge starts to spread inside bubbles already during the unfolding of the epidemic. So, even if complete knowledge spreading remains quite slow, pieces of information that circulate inside bubbles guarantee that part of the knowledge is acquired from the beginning (in the reported case, already at day 59). This important achievement still has to pay the price of (i) a high number of simultaneously active infected individuals, and (ii) a still long time before all the pieces of knowledge are able to reach all the nodes.\\
% Bubbles with tournaments
Hence, we consider an additional containment strategy: we add to the node clustering of nodes, represented by the social bubbles, a temporal clustering, thus obtaining temporal social bubbles. We leverage again the value of $p$ setting the ratio of intra-bubbles' and inter-bubbles' connections but, in this strategy, the inter-bubbles' connections, instead of involving nodes of  random different bubbles, are now concentrated only between specific couples of bubbles (as in the toy example depicted in Fig.~\ref{fig_time_evol}D). So, each bubble (e.g. a team or department within an organization) only interacts with another bubble (e.g. a different team or department) at a specific time. Then, with a time periodicity $d$ the couples change in such a way that, at the end of the simulation, each bubble has interacted with each other bubble. The result is that knowledge starts to grow from the early stages of the simulations inside each couple of bubbles and each node of one bubble easily acquires the ideas of its matched bubble, doubling the nodes' average knowledge. In a similar manner, once matches are updated, nodes are able to acquire knowledge from another bubble, and gradually augment their knowledge following a staggered growth. In Fig.~\ref{fig_time_evol}D, this dynamic is clearly visible: the purple line shows the process of knowledge acquisition, which grows more rapidly in correspondence with the update of bubble matches (every $d=10$ days in the reported simulations), and slows down once most of the matched bubble nodes have acquired the new piece of knowledge. 
For what concerns the disease spreading, leveraging the peculiar structure of node interactions, it naturally remains confined between a limited number of bubbles and, as a consequence, the number of infected individuals grows more slowly. Moreover, the number of simultaneously active infected does not reach a high value since a fraction of the infected nodes can recover within the time that matches between bubbles are changed. By observing the curves in Fig.~\ref{fig_time_evol}D we notice that the disease curve and the knowledge one are partially overlapping, indicating that, in this framework, the epidemic can coexist with the diffusion of knowledge by social face-to-face interactions.\\
% Optimal p
In Fig.~\ref{fig_opt_p_10cl} we report, for different strategies and different parameters' settings, four significant quantities characterizing disease and knowledge spreading: (i) the maximum number of simultaneously active infected individuals (orange horizontal dashed lines in Fig.~\ref{fig_time_evol}), (ii) the time at which the average knowledge acquired by nodes becomes equivalent to the number of different pieces of knowledge initially present in one cluster, (iii) the time at which it reaches 50\% of the total knowledge, and (iv) the time at which it reaches 80\% of the total knowledge (the three purple vertical lines in Fig.~\ref{fig_time_evol}).
For the quarantine strategy, these indicators are reported versus $\varepsilon_T$ and while, clearly, infected individuals decrease with $\varepsilon_T$, the knowledge times it is only marginally affected by it. For the bubbles' strategy, the indicators are instead depicted as functions of $p$, where increasing $p$ means making the bubbles more self-contained. In particular, we consider the case with temporal clustering for three different values of $d$ (i.e. 5, 10, and 20 days), and without temporal clustering. In all cases, we notice that the number of infected individuals decreases with $p$, not drastically as for $\varepsilon_T$, but significantly. This is due to the fact that closer bubbles tend to maintain the disease confined to a few bubbles, while the other nodes remain safe. The lowest numbers of infected individuals are obtained with temporal clustering of 20 days, in fact with longer temporal clustering there is more chance that infected individuals inside a bubble recover or are quarantined before they have the possibility to meet new susceptible nodes. For what concerns knowledge times, instead, we notice a peculiar behaviour with respect to the modularity, showing in all cases a minimum value in $p$, corresponding to an optimal value which varies according to different conditions. We also notice that in order to shorten times to reach partial knowledge, longer temporal clustering are to be preferred (see dark purple curves), while to shorten total knowledge times we need to decrease the length of the tournament $d$ (see light purple curves).

%%%%%%%%%%%%%%%%%%%%%%%%%%%%%%%%%%%%%%%%%%%

%%%%%%%%%%%%%%%%%%%%%%%%%%%%%%%%%%%%%%%%%%%
%%%%%%%%%%%%%% DISCUSSION %%%%%%%%%%%%%%%%%
\section{Discussion}
In this work, we have investigated the social bubbles' framework as a potential strategy for controlling disease spreading and simultaneously allowing higher levels of face-to-face interactions, which in turn facilitate the process of knowledge diffusion. More specifically, we have considered different settings of the social interactions' network organization in bubbles with the aim of finding an alternative strategy to preventive quarantines.

%Comparing the results from the different frameworks, it is evident that, from the point of view of reducing the number of infected individuals, the quarantine strategy is the most effective one. 
From the point of view of reducing the number of infected individuals, the quarantine strategy is the most effective one. However, arguably, quarantines impede social contacts and, as a consequence, all the diffusion processes that do not rely on simple contagion models but require multiple contacts to spread, as is the case for the considered knowledge-spreading process. The bubbles' strategy instead permits face-to-face interactions between people, albeit limited to smaller groups than to an entire organization, school, university campus, workplace, etc. These free physical interactions allow complex dynamics (like complex contagions) to emerge, as it is demonstrated by the progress of knowledge spreading, here represented by the threshold process. We notice in fact that, despite the acquirement of the entire knowledge set is still reached after a long time, the singular ideas are free to circulate since the beginning inside the bubbles and are then easily acquired by the nodes (see Fig.~\ref{fig_time_evol}C).\\
% Melinda Mills
%The idea of this work stemmed from the article by Block et al.~\cite{block2020social}, which proposes a solution to flatten the curve of infection by modifying the network structure to increase modularity. Their strategy however consists in cutting inter-bubble links and not in rewiring them.
Other strategies to flatten the infection curve by making use of social bubbles consist in cutting inter-bubble links instead of rewiring them~\cite{block2020social}.
However, the reduction of connections incontestably induces a slowdown of the epidemic spreading, which is not entirely due to the organization in bubbles. In our work, we show instead that the bubbles strategy is effective even if the number of links is maintained constant. In particular, we use networks with different levels of modularity, quantified by different values of $p$, but we generate the networks always with the same average number of links.%, so changing $p$ is equivalent to rewiring links.

% threshold model, communities, optimal p
One of the most salient results of this work is that the existence of communities in a network facilitates knowledge diffusion (i.e. complex contagion), limiting collateral confinement, while it mitigates disease diffusion (i.e. simple contagion). An analogous conclusion about knowledge diffusion, implemented as a linear threshold model, has been obtained by Nematzadeh et al.~\cite{nematzadeh2014optimal} (and by Peng et al.~\cite{peng2020network} in a follow-up of the first work). They observe that strong communities enhance local spreading, while weak communities enhance global spreading, and they find an optimal range of intermediate values for community strength (analogous to our $p$) that maximize diffusion and the speed of cascades. This result agrees with our results on knowledge spreading; %despite a slightly different model and evaluation approach. 
%Their approach is a bit different, they consider a linear threshold model and they look at the stationary solution of the process, 
in fact, with our analyses we also observe the existence of an optimal value of $p$ corresponding to the shortest times of knowledge diffusion (see Fig. 3). The minimum in $p$  tells us that the best performances of the bubbles' strategy are not monotonic with $p$, on the contrary, very high or very low values of $p$ slow down the knowledge diffusion. In fact,  trivially, if $p$ is too small we lose the effect of population partition and what we observe is a mixed population without bubbles and without quarantines, so a very inefficient network structure. If, instead, we increase $p$, i.e. we make the bubbles progressively more self-contained, we interestingly observe progressively longer knowledge spreading times, suggesting that a certain level of promiscuity between bubbles is instead advisable. 
% p does not change
It is important to notice that the existence of quarantines and isolations does not affect the ratio $p$ between the amount of intra- and inter-bubble connections. In Supplementary Material, Fig.~\ref{fig_eff_p}, we show indeed that, while the number of overall connections in the network is significantly reduced during the central phase of the disease epidemic, both within and between clusters, the ratio $p_{intra}/p_{inter}$ remains stable around the value $p$ set to generate the networks. This confirms the validity of the parameter $p$ to discern the different networks and the resulting processes.\\
% tournaments
The greatest advantage is however obtained with the insertion of bubbles' temporal clustering. In fact in that case we have a combined effect: the existence of bubbles allows to keep the epidemics under control, and the temporal clustering enforce the nodes to be exposed to periodically different ideas. The result is a faster acquisition of knowledge while the disease is kept under control. This strategy reveals impressively effective and, up to our knowledge, it is a completely novel idea.\\
% bubbles sizes
All these results are confirmed by considering different sizes of bubbles (with the same number of nodes and the same average number of links), in particular 5 bubbles of 136 individuals and 20 bubbles of 34 individuals, as reported in Section S2 of the Supplementary Material.

Our work comes with some limitations that could be addressed by further exploring this research direction in the future. %For instance, we are only considering synthetic networks, a necessity stemmed from the will to tune the networks' structure without many constraints, but it would be interesting to investigate real networks too.
First of all, we are only considering synthetic networks, which are generated with random interactions without temporal correlations, clustering, or other structural information that would make them more similar to real networks of interactions. 
The reason for this choice stemmed from the need to investigate the effect of social bubbles, disentangled from other possible structural constraints that could affect the dynamics. The choice of random networks ensures that the only structure existing in the considered graphs is the modularity, which we introduce and control by setting the parameter $p$ when generating the networks.
This allows us to directly scrutinize the phenomenon and draw untwisted conclusions. Envisaging possible applications to real networks will be matter of future investigations.
Moreover, the effect of network density is not explored, but we test different network sizes (increasing or decreasing the number of nodes and changing consequently the number of links so as to keep density fixed) and different bubble sizes and we observe that the results essentially do not change (see Section S3 of the Supplementary Material). Finally, only some of the simple and complex spreading parameters are explored, which implies that the results about optimal $p$ are not general, they only apply in this specific context. However, the interesting result is that a minimum exists, even if its exact value will probably change by changing the parameters.\\
In conclusion, we realize that the temporal social bubbles strategy represents a valid alternative to other NPIs like preventive quarantines when looking for a solution permitting to coexist with a spreading disease. This strategy, while still affecting the social structure of interactions, allows the pursuit of a series of otherwise hardly attainable collective goals that prove fundamental to the growth and social enrichment of society and individuals, from collaborations to social relationships, from knowledge transfer to opinion exchange. These results could spur innovative approaches to epidemic control strategies, for example, based on different interaction-mixing prescriptions in different settings, such as home, work and leisure places. While a systematic assessment of mixed strategy approaches is needed to better inform policymakers, our results provide a solid proof of concept for social bubble strategies, proving its efficacy under the prescription of temporal clustering, evidencing how the choice of a strategy solely focused on flattening the curve can dramatically affect knowledge diffusion and social cohesion.

\section{Methods}

	\subsection{Disease spreading}
	\label{sec_dis_spread}

	The disease-spreading model is inspired by previous literature on COVID-19 models~\cite{ferretti2020quantifying, he2020temporal,cencetti2021digital}.
	Each numerical simulation starts with one random infected individual, who has been infected for a number of days, $\tau$, randomly sampled between 0 and 10. The variable $\tau$ is important in the spreading process since we assume that infectiousness, i.e. the probability of transmitting the disease, of an infected individual depends on the time since their own contagion, with a function $\omega(\tau)$ which has a maximum peak at around 5 days (see Fig.~\ref{fig_omega_s} in the Supplementary Material). Such probability governs which individuals, among those that the infected seed meets according to the temporal network, will contract the disease.  These can in turn infect their contacts. We assume that every infected individual becomes recovered, hence immune\footnote{If we extend the time span of the simulations we should consider the fact that immunity only exists for a finite period of time, however, since we are considering only around 4 months, considering that recovered people remain immune until the end of the simulations is a good approximation of reality.}, after 25 days~\cite{barman2020covid}. We assume that 80\% of infected individuals become symptomatic after being infected, with a symptom onset probability which increases in time according to a function $s(\tau)$ \cite{cencetti2021digital} (see Fig.~\ref{fig_omega_s} in the Supplementary Material). As soon as they show symptoms they have a probability  $\varepsilon_I$ of being isolated. If this does not happen they go on spreading, otherwise their contacts are cut for the next 30 days, at the end of which they will become recovered. When an individual is isolated, their past contacts (last 7 days) are traced and preventively quarantined with a probability $\varepsilon_T$. The quarantined individuals can be infected (true positive) or susceptible (false positive). In the first case, if they show symptoms during quarantine, they will become isolated (the only difference with quarantine is that their past contacts are traced and, in the end, they will be recovered), otherwise, they finish the quarantine after 10 days and  they are released.

	\subsection{Network generation}
	
We generate a synthetic temporal network with $N$ nodes organized in $N_c$  clusters, characterized by a strength of network modularity $p=p_{intra}/p_{inter}$, and a total number of  links $L$. We hence generate a series of static networks which are going to constitute the layers of the temporal network. In each layer, nodes are classified into clusters (the first $n_1$ nodes in the first cluster, the second $n_2$ nodes in the second cluster, and so on, with $n_i$ the a priori chosen number of nodes in cluster $i$, the same for each static network). All the temporal layers are characterized by the same values of intra-cluster connections probability $p_{intra}$ and of inter-cluster connections probability $p_{inter}$. 

The static networks are generated using the Stochastic Block Model~\cite{holland1983stochastic}, once all the parameters ($N_c$, $n_i$, $p_{intra}$, $p_{inter}$) have been fixed.

Since $p_{intra}$ and $p_{inter}$ are not independent of each other but are constrained by the value of $L$, we need first to find the function that associates these three variables. We consider the  general case where clusters do not contain the same number of nodes, but each cluster $i$ contains a number $n_i$ of nodes. This means that, statistically, in cluster $i$ we can find a number of internal links given by:
\begin{equation}
l_{intra}^i = \frac{n_i(n_i-1)}{2} p_{intra}
\end{equation}
and summing over all clusters we obtain:
\begin{equation}
l_{intra} = \sum_{i=i}^{N_c} l_{intra}^i  = \sum_{i=i}^{N_c}  \frac{n_i(n_i-1)}{2} p_{intra}.
\end{equation}
For what concerns the inter-cluster links, instead, we should consider that each node in cluster $i$ can be connected to a number of external nodes
$(N-n_i) p_{inter}$. 
This is true for every node in the cluster, so it should be multiplied by $n_i$ to find the number of connections of cluster $i$ that point to other clusters:
\begin{equation}
l_{inter}^i = n_i(N-n_i) p_{inter}.
\end{equation}
To obtain the total number of inter-cluster links we just have to sum over all the clusters and divide by 2, to avoid double counting:
\begin{equation}
l_{inter} = \sum_{i=i}^{N_c} \frac{l_{inter}^i}{2} = \sum_{i=i}^{N_c} \frac{n_i(N-n_i)}{2} p_{inter}.
\end{equation}
The total number of links can therefore be written as:
\begin{equation}
L = l_{intra} + l_{inter} =  \sum_{i=i}^{N_c} [\frac{n_i(n_i-1)}{2} p_{intra} + \frac{n_i(N-n_i)}{2} p_{inter}].
\end{equation}
This equation represents the constraint between $L$, $p_{intra}$, and $p_{inter}$. 
In our case we consider all the clusters with the same number of nodes: $n_i = N/N_c \equiv n_c$ $\forall i$, hence it reduces to 
\begin{equation}
L = N_c [\frac{n_c(n_c-1)}{2} p_{intra} + \frac{n_c(N-n_c)}{2} p_{inter}].
\label{eq_L}
\end{equation}
The networks that we generate for this manuscript have a fixed number of links, $L=400$. By varying the chosen value for $p_{intra}$ we can obtain different choices of $p_{inter}$ (and hence of $p$), always maintaining constant the number of links $L$, by inverting equation~\ref{eq_L}:
\begin{equation}
p_{inter} = \frac{2L}{N_c n_c(N-n_c)} - \frac{n_c-1}{N-n_c} p_{intra}.
\end{equation}

\section{Code availability}

The code used for the generation of temporal network, simulations and analysis is available at:
\url{https://github.com/giuliacencetti/Social_bubbles}

\section{Competing Interests}
The authors declare no competing interests.

\section{Acknowledgements}
G.C., G.S., and B.L. acknowledge the support of the PNRR ICSC National Research Centre for High Performance Computing, Big Data and Quantum Computing (CN00000013), under the NRRP MUR program funded by the NextGenerationEU. L.L. has been supported by the ERC project ``IMMUNE'' (Grant agreement ID: 101003183).

	%\addcontentsline{toc}{chapter}{Bibliography}
	\bibliographystyle{ieeetr} 
	\bibliography{biblio}

\clearpage

\newpage

%\appendix
\input{supplementary}

\end{document}

%% file: supplementary.tex
\pagebreak
\widetext
\begin{center}
\textbf{\large Supplementary material for Temporal clustering of social interactions trades-off disease spreading and knowledge diffusion}
\end{center}

\gdef\thefigure{\arabic{figure}}
\gdef\theequation{\arabic{equation}}
\gdef\thetable{\arabic{table}}
\setcounter{figure}{0}
\setcounter{equation}{0}
\setcounter{table}{0}
\setcounter{section}{0}

\renewcommand{\thefigure}{S\arabic{figure}}
\renewcommand{\thesection}{S\arabic{section}}
\renewcommand{\thetable}{S\arabic{table}}

\section{Details of disease spreading}

\begin{figure}[h!]
	\centering
	\includegraphics[width=0.3\textwidth]{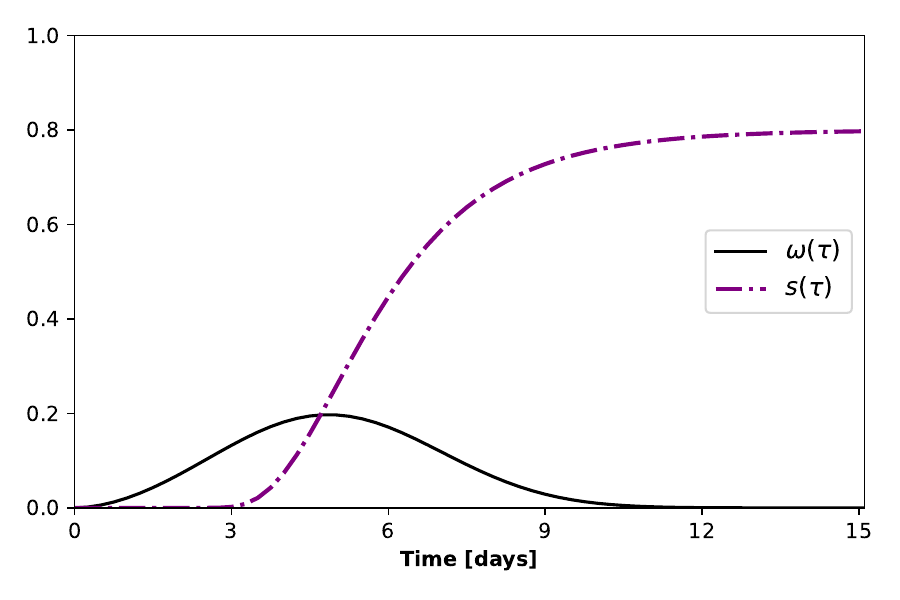}
 	\caption{Infectiousness $\omega(\tau)$ and cumulative distribution of onset times $s(\tau)$.}
 	\label{fig_omega_s}
\end{figure}

Fig.~\ref{fig_omega_s} shows in black $\omega(\tau)$, the probability of an infectious individual transmitting the disease as a function of the time since their own contagion, $\tau$. The function has a maximum around 5 days after the contagion occurred. 
The purple curve instead represents $s(\tau)$, the cumulative distribution probability of symptom onset as a function of $\tau$.
Both functions have been used in \cite{cencetti2021digital}, inspired by \cite{ferretti2020quantifying}.

\clearpage %%%%%%%%%%%%%%%%%%%%%%%%%%%%% %%%%%%%%%%%%%%%%%%%%%%%%%%%%%

\section{Results with different quantities and sizes of bubbles}

\begin{figure*}[h!]
\includegraphics[width=\textwidth]{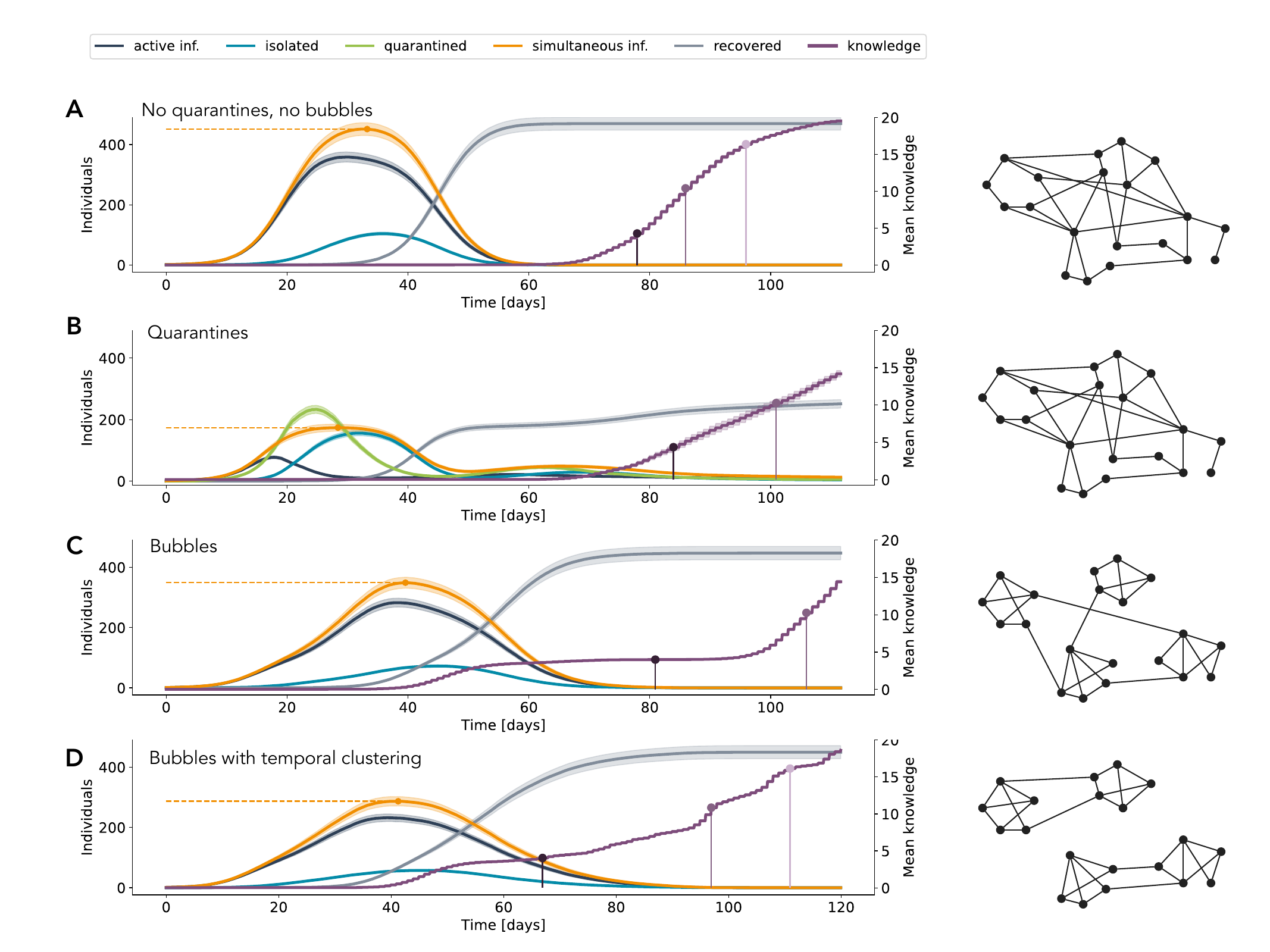}
\caption{Time evolution of disease and knowledge spreading, results of 200 simulations on temporal networks organized in 5 bubbles with 136 nodes. A) $p=5$ (i.e. connections inside and outside bubbles are of the same order, in practice, bubbles do not exist), $\varepsilon_T = 0$ (i.e. no quarantines). B)  $p=5$, $\varepsilon_T = 0.1$ (i.e. quarantines without bubbles). C) $p=199$, $\varepsilon_T = 0$ (i.e. bubbles without quarantines). D) $p=199$, $\varepsilon_T = 0$, plus temporal clustering of 10 days.
The parameter $\varepsilon_I = 0.1$ for all these cases.}
\label{fig_time_evol_5cl}
\end{figure*}

\begin{figure*}[h!]
\includegraphics[width=\textwidth]{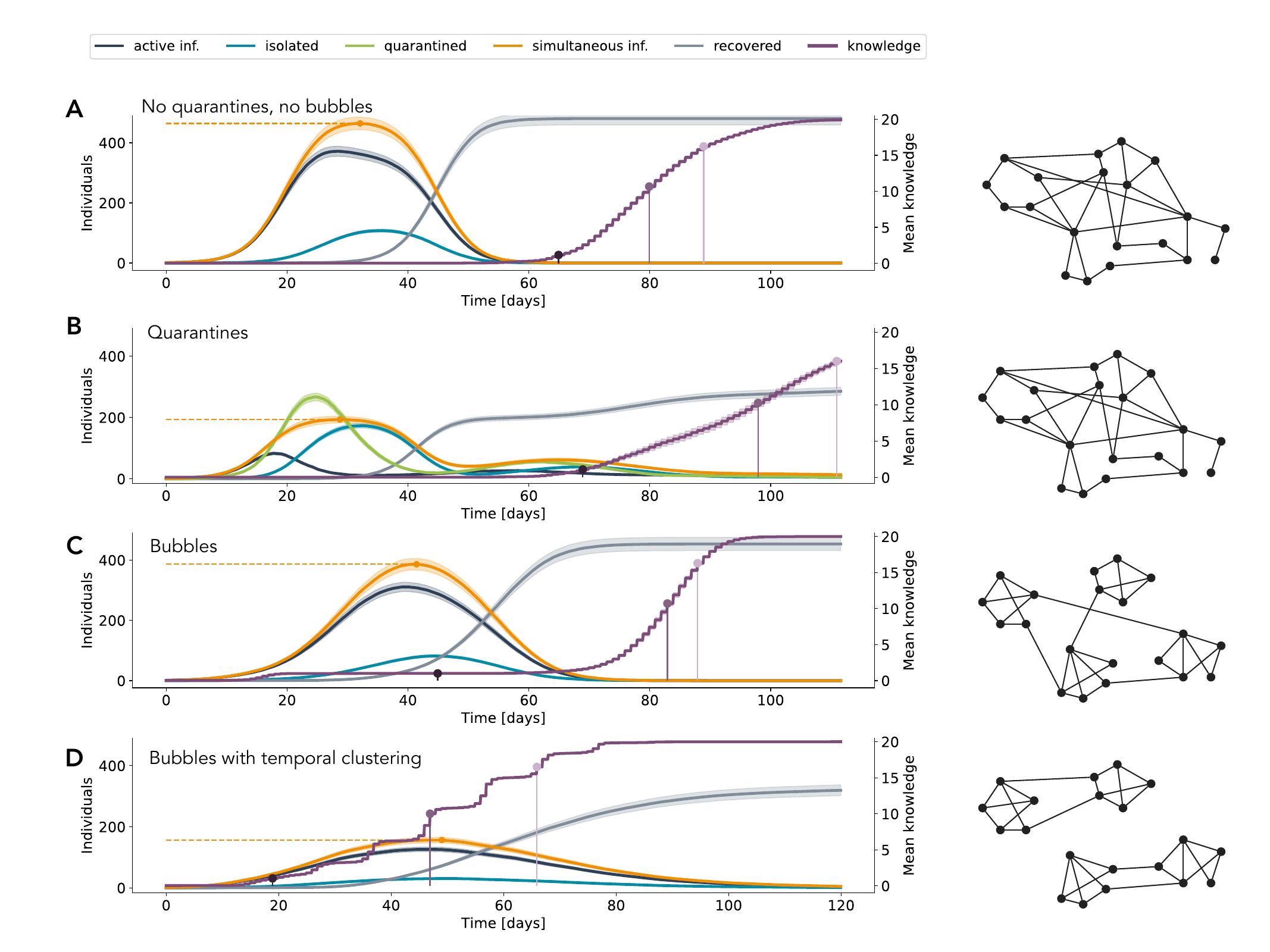}
\caption{Time evolution of disease and knowledge spreading, results of 200 simulations on temporal networks organized in 20 bubbles with 34 nodes.}
\label{fig_time_evol_20cl}
\end{figure*}

\begin{figure*}[h!]
\includegraphics[width=\textwidth]{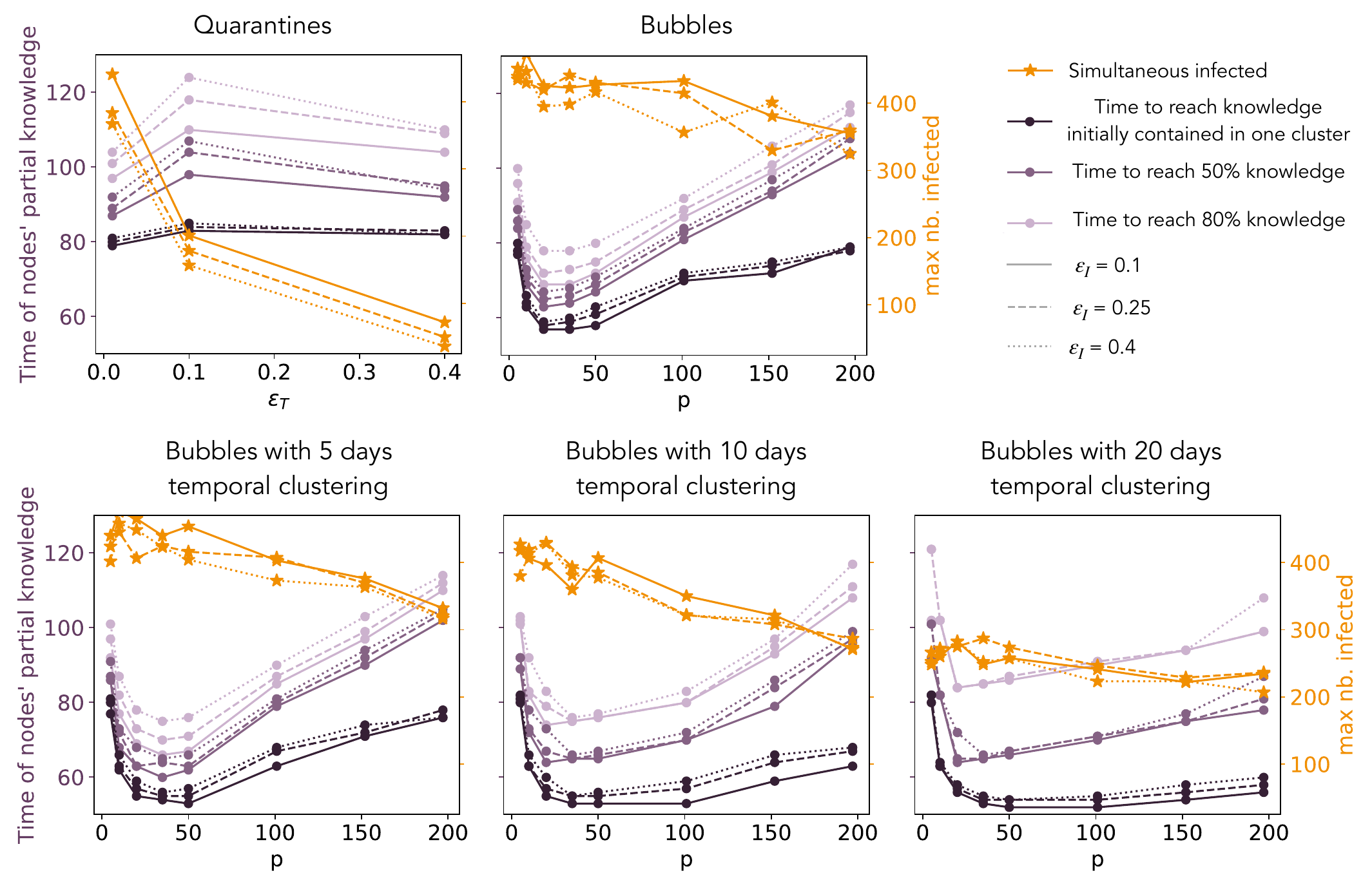}
\caption{5 clusters. Orange stars represent the max number of infected. The other symbols represent the time at which the nodes reach, on average, 20\%, 50\%, and 80\% of knowledge. The first percentage corresponds to acquiring the entire knowledge initially contained in one cluster from all the nodes. We report results for $\varepsilon_I = 0.1$ (continuous lines), $\varepsilon_I = 0.25$ (dashed lines), $\varepsilon_I = 0.4$ (dotted lines).}
\label{fig_opt_p_5cl}
\end{figure*}

\begin{figure*}[h!]
\includegraphics[width=\textwidth]{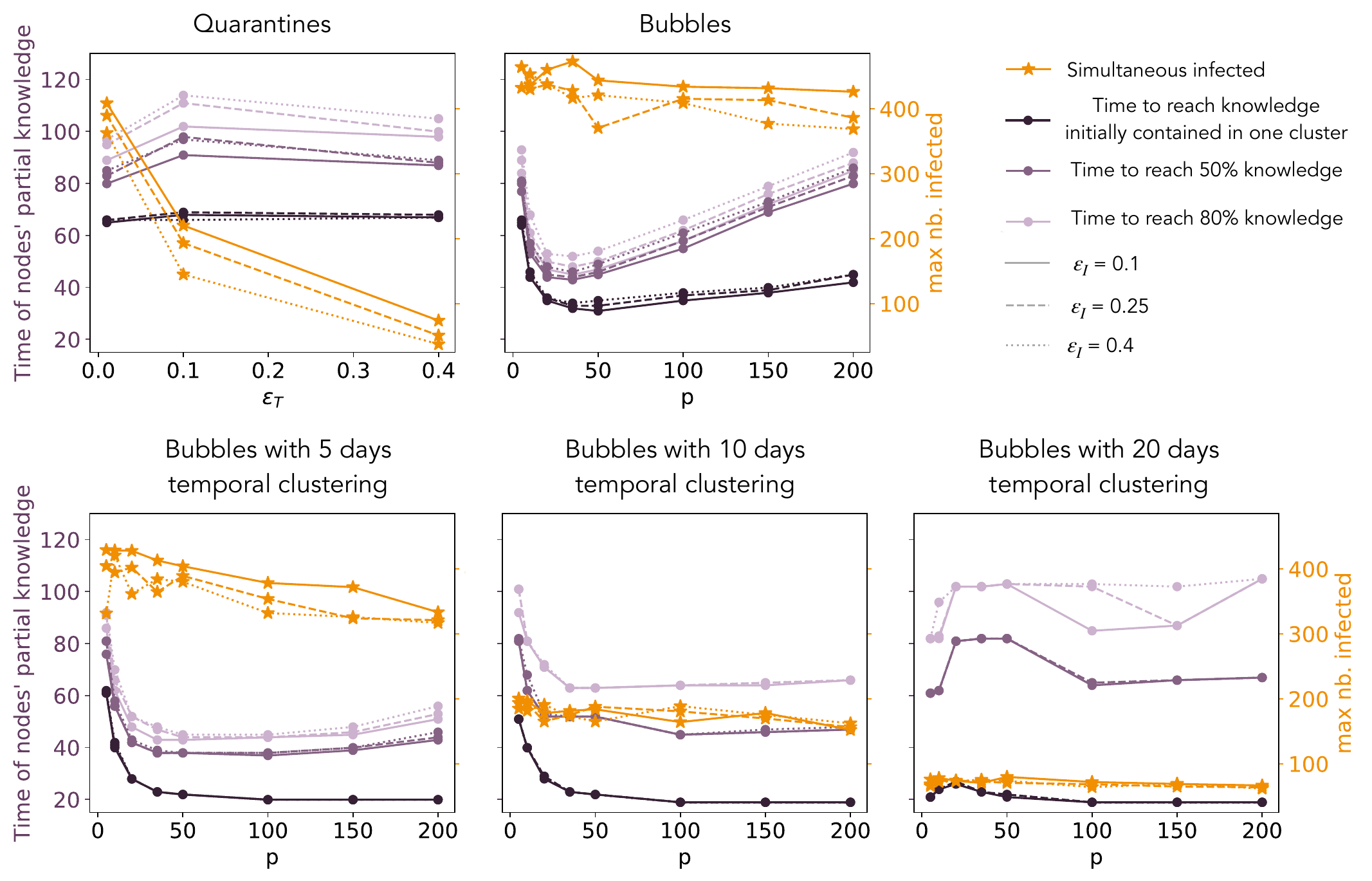}
\caption{20 clusters. Parameters are set as in Fig.~\ref{fig_opt_p_5cl}}
\label{fig_opt_p_20cl}
\end{figure*}

In the main text, we show the results obtained when the network with 680 nodes is partitioned into 10 clusters of 68 nodes (with several levels of modularity). Here we investigate two different partitions: 5 clusters of 136 nodes and 20 clusters of 34 nodes. Figs.~\ref{fig_time_evol_5cl}, \ref{fig_time_evol_20cl} represent the analogous of Fig. 2 of the Main Text and Figs.~\ref{fig_opt_p_5cl}, \ref{fig_opt_p_20cl}  the analogous of Fig. 3 of the Main Text. The results are similar to those obtained for 10 bubbles, suggesting that the number and size of bubbles have a lower effect on the dynamical processes than the modularity $p$.

\clearpage 

%%%%%%%%%%%%%%%%%%%%%%%%%%%%%

\section{Results with different numbers of nodes}

\begin{figure*}[h!]
\includegraphics[width=\textwidth]{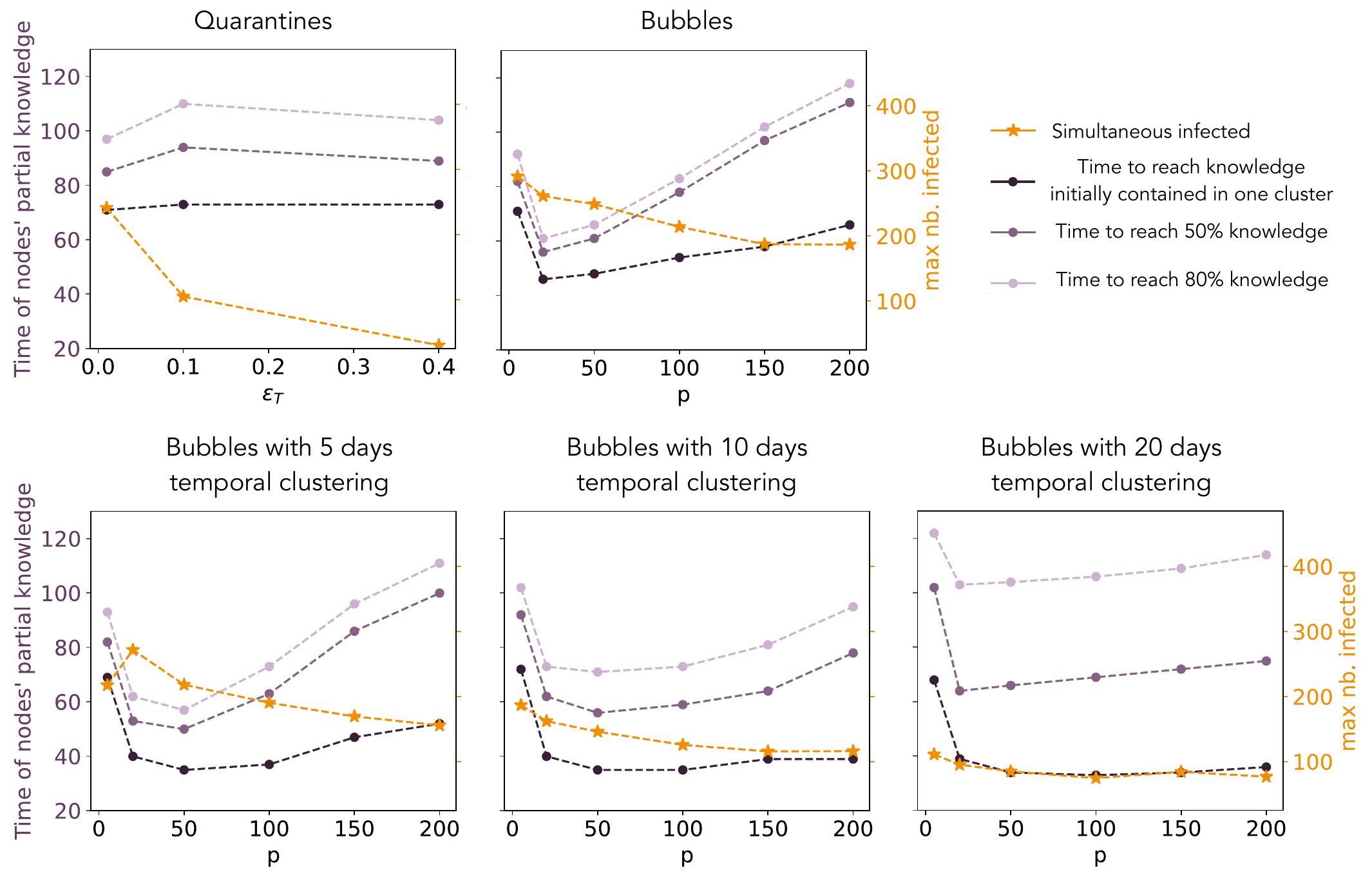}
\caption{Results obtained with a smaller temporal network: 510 nodes}
\label{fig_opt_p_510nodes}
\end{figure*}

\begin{figure*}[h!]
\includegraphics[width=\textwidth]{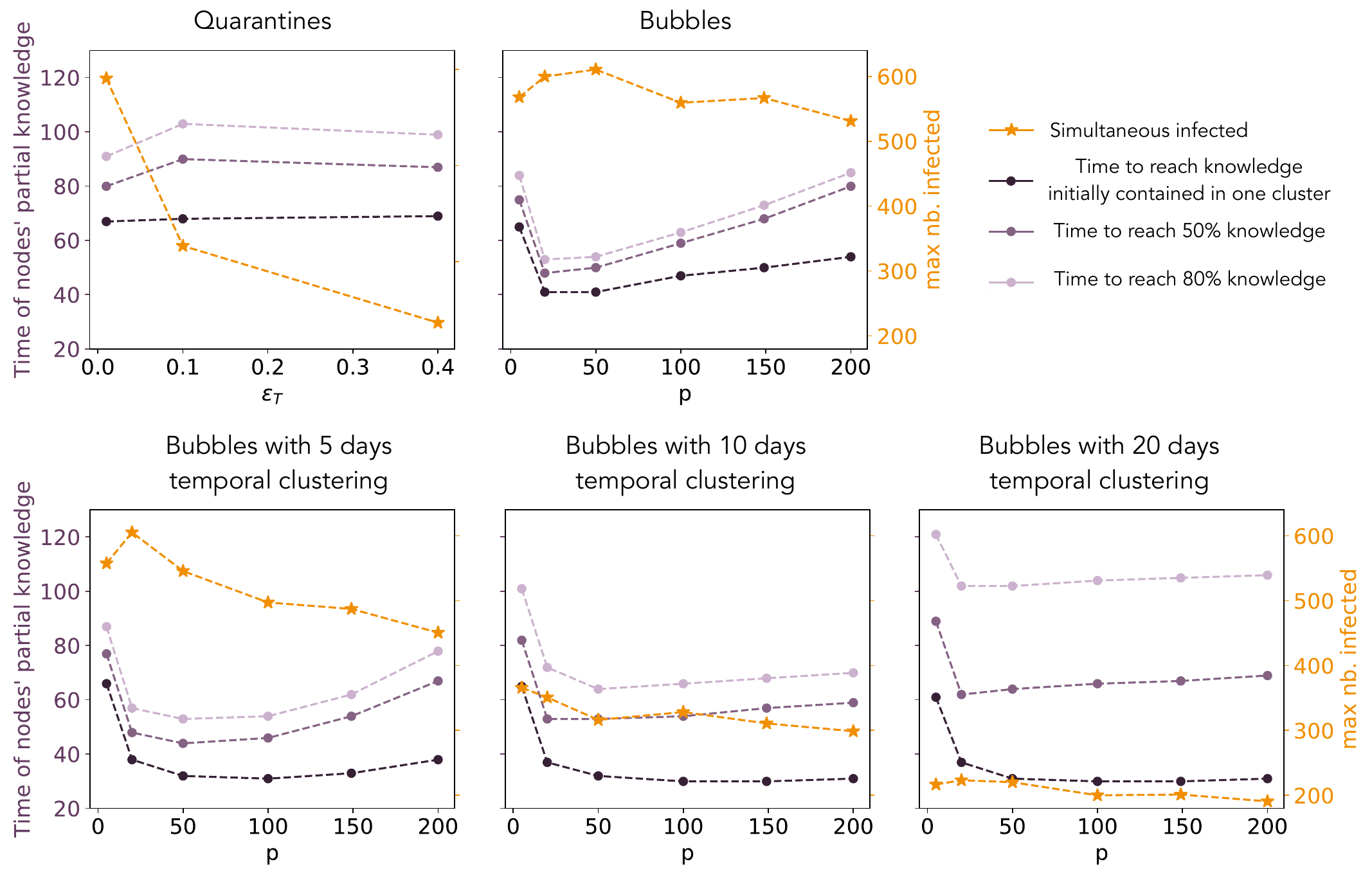}
\caption{Results obtained with a larger temporal network: 850 nodes}
\label{fig_opt_p_850nodes}
\end{figure*}

For the sake of completeness, we report results obtained for different numbers of nodes with respect to the 680 nodes of the Main Text, in particular a reduction of 25\%, corresponding to 510 nodes, and an increase of 25\%, corresponding to 850 nodes. The number of links is changed accordingly so as to keep density fixed (225 links on average for the networks of 510 nodes, 400 links for 680 nodes, 625 links for 850 nodes). Results are depicted in Figs.~\ref{fig_opt_p_510nodes} and \ref{fig_opt_p_850nodes} respectively. Comparing them with Fig.~\ref{fig_opt_p_10cl} we notice that changing the size of the network does not change the conclusions obtained in the Main Text.

\clearpage %%%%%%%%%%%%%%%%%%%%%%%%%%%%%

\section{Alternative visualization of results}

\begin{figure*}[h!]
\includegraphics[width=\textwidth]{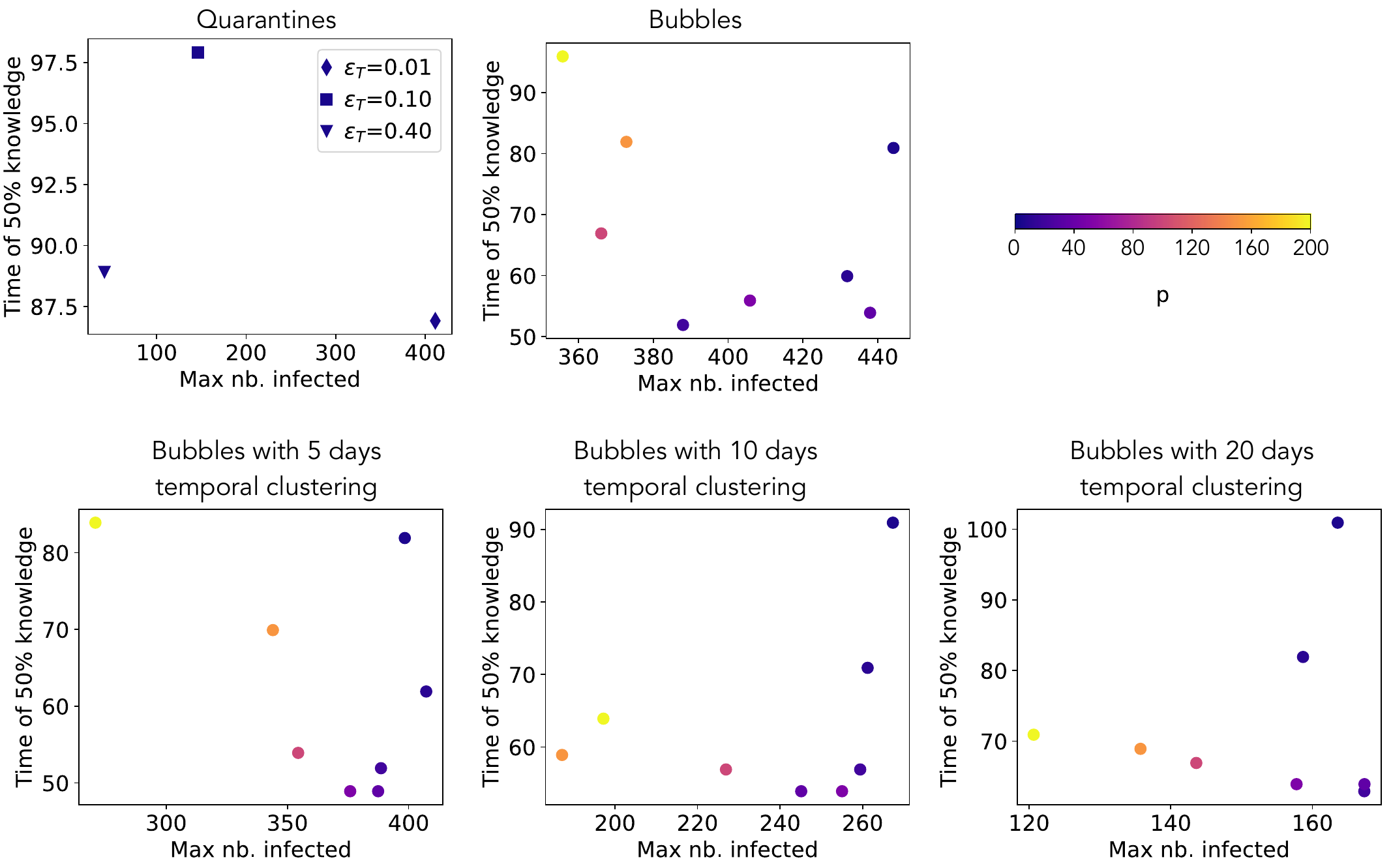}
\caption{Time to reach 50\% knowledge and number of maximum infected for simulations with 680 nodes, 10 clusters, for $\varepsilon_I$=0.5. These results are also reported in Fig.~\ref{fig_opt_p_10cl}, here are shown as knowledge time vs. infected. In the panel with quarantines, different symbols correspond to different values of $\varepsilon_T$, in the other panels different colours correspond to different values of modularity $p$, while $\varepsilon_T=0$. }
\label{fig_opt_p_10cl_pareto}
\end{figure*}

\clearpage %%%%%%%%%%%%%%%%%%%%%%%%%%%%%

\section{Stability in time of effective $p$}

\begin{figure*}[h!]
\includegraphics[width=\textwidth]{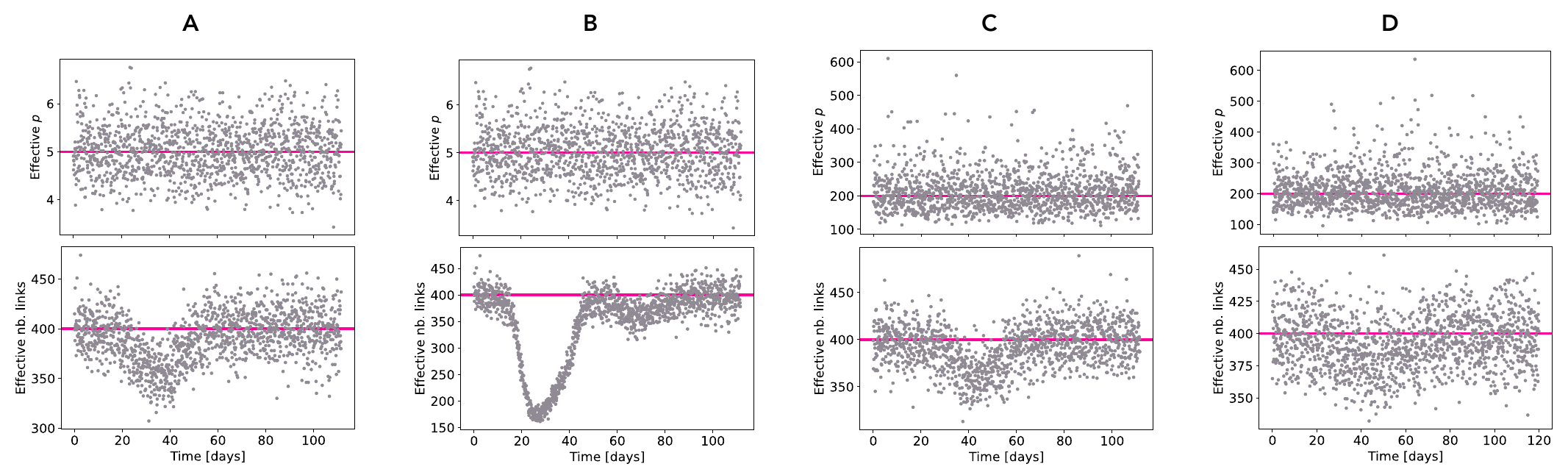}
\caption{The effective value of $p$ is computed on the networks where links of isolated nodes have been cut and are reported in grey in the upper panels (each dot corresponds to one  layer of the temporal network). The horizontal line reports the value of $p$ set to generate the networks. The lower panels report the number of total connections in the same networks, while the horizontal line shows the number of connections statistically assigned when networks are generated. We observe a drastic reduction of links during the epidemic phase, not mirrored by any change in the effective $p$ which remains stable. The networks used for these figures are the same used for the simulations in the main text, with 10 bubbles of 68 nodes, $\varepsilon_I=0.25$, $\varepsilon_T=0$ and three values of $p$: $5$, $100$, and $200$.}
\label{fig_eff_p}
\end{figure*}

The parameter $p$ proves fundamental to explain the results shown in the Main Text. This parameter is set to generate the layers of the temporal networks and establishes the ratio between the probability of the existence of links within clusters and the probability of the existence of links between clusters, $p=p_{intra}/p_{inter}$. However, the process involves isolations and quarantines of nodes, which implies that all the links of the corresponding nodes are cut in the temporal layers corresponding to their isolation/quarantine period. This causes a general reduction of connections in the network during the epidemic phase, as shown in the lower panels of Fig.~\ref{fig_eff_p}. One could therefore wonder if the ratio between intra and inter-cluster links is preserved or altered by these modifications, where this last possibility would question the reliability of parameter $p$ in characterizing the processes under study. We hence computed the number of connections inside and outside bubbles in the effective networks, i.e. those without the links of isolated and quarantined nodes, obtaining an effective value of the ratio $p$. This is reported in the upper panels of Fig.~\ref{fig_eff_p} for the four different cases (A, B, C, D) represented in Fig.~\ref{fig_time_evol} of the main text. The effective $p$ fluctuates  around the value that has been set to generate the networks (represented by the horizontal line) without any significant change in time, in contrast to the number of total links which shows a dramatical change. This means that, while both intra and inter-cluster links are reduced for an interval of time, their ratio remains constant, thus confirming the reliability of $p$ in identifying networks and discerning the resulting processes.

\clearpage 
%%%%%%%%%%%%%%%%%%%%%%%%%%%%%

\section{Collateral confinement of the quarantine control strategy}
\label{sec_confinement}
In terms of impact on individuals' lives, quarantines (i.e. preemptive confinement strategy) are probably one of the policies with the highest social costs for individuals. Jointly with high-quality tracing and testing policies, it can significantly reduce infective disease spread. In this work, we use this strategy as a reference for comparison with the social bubble framework, whose social costs are arguably lower. In the main manuscript, we compare quarantines and social bubbles with temporal clustering to understand which strategy has a smaller effect in slowing down the knowledge diffusion process. As a reference, since a direct comparison with the temporal clustering bubbles is not possible, we report, for different simulation configurations, the percentage of the population which was confined without being infected as an additional measure of the collateral costs.
\begin{table}[hb!]
\centering
    \begin{tabular}{l|ccc}
    % \toprule
    & $\epsilon_T$ = 0.01 & $\epsilon_T$ = 0.1 & $\epsilon_T$ = 0.4 \\
    \hline
    % \midrule
    $\epsilon_I$= 0.1  & 1,9\%& 30,6\%& 53,1\%\\
    $\epsilon_I$= 0.25 & 4,6\%& 42,2\%& 51,3\%\\
    $\epsilon_I$= 0.4  & 6,9\%& 41,9\%& 59,6\%\\
    % \bottomrule
    \end{tabular}
\caption{\emph{Percentage of non-infected quarantined population.} For each configuration of the testing ($\epsilon_I$) and quarantine ($\epsilon_T$) parameters, we report the fraction of the population that was quarantined at least once without being infected.}
\end{table}